\documentclass[epj,onecolumn]{webofc}
\usepackage[varg]{txfonts}   
\woctitle{KLOE-2 Workshop on $e^+e^-$ collider physics at 1 GeV}
\begin{document}
\title{Latest studies on the reaction $e^+ e^- \rightarrow K^+ K^- \gamma$}
%
%

\author{\firstname{Leonard} \lastname{Le\'sniak}\inst{1}\fnsep\thanks{\email{Leonard.Lesniak@ifj.edu.pl}} \and
        \firstname{Micha\l{}} \lastname{Silarski}\inst{1}\fnsep\thanks{\email{Michal.Silarski@uj.edu.pl
        }} 
}

\institute{Institute of Physics, Jagiellonian University, 30-348 Krak\'ow, Poland 
          }

\abstract{Recent theoretical studies of the $e^+ e^- \rightarrow K^+ K^- \gamma$ process are described.
Three main reaction mechanisms are considered: the initial state radiation, the final state radiation and the strong interaction between the outgoing $K^+ K^-$ mesons. The $K^+ K^-$ effective mass distributions are derived
for three different models which in past have been used for a description of the $e^+ e^- \rightarrow \pi^+ \pi^- \gamma$ data. Also the numerical results for the angular photon and kaon distributions are presented.
A new model of the $e^+ e^- \rightarrow M_1 M_2 \gamma$ reactions is outlined which can serve for multichannel
analyses of the radiative processes with a production of two pseudoscalar mesons $M_1$ and $M_2$. 
 }
\maketitle
\section{Introduction}
\label{intro}

Strong interactions of the strange mesons $K^+$ and $K^-$at low energies are not well known.
Since the $K^+ K^-$ threshold lies quite close to the DA$\Phi$NE accelerator energy,
the low energy $K^+ K^-$ interactions can be studied using  data collected by the KLOE experiment.
Masses of the scalar resonances $f_0(980)$ and $a_0(980)$ are close to 1 GeV.
There are, however, large uncertainties in their values. 
According to the Particle Data Group
estimations \cite{PDG} the $f_0(980)$ mass equals to 990 $\pm$ 20 MeV and its width lies between 10
and 100 MeV, while the $a_0(980)$ mass is 980 $\pm$ 20 MeV and the width value varies between 50 
and 100 MeV. 
Let us stress here that the parameters of the scalar resonances found in experimental analyses are very much model dependent. 

As an example let us consider the reaction $e^+ e^- \rightarrow \pi^+ \pi^- \gamma$ studied by the KLOE Collaboration in 2006 \cite{KLOE2006}. 
Fits to the experimental data done using the following two models: the kaon-loop model \cite{KL} and the so-called no-structure model \cite{NS}, gave quite different values of the $f_0(980)$ mass ranges: 980-987 MeV
in the first case and 973-981 MeV in the second case. 
The intervals of the maximal variations of the $f_0(980)$ coupling constants to $K^+ K^-$ were:
5.0-6.3 GeV for the first model and  1.6-2.3 GeV for the second one.
The corresponding numbers for the $f_0(980)$ coupling constants to $\pi^+ \pi^-$ were:
3.0-4.2 GeV and 0.9-1.1 GeV, respectively.
One can attribute the differences between the resonance parameters to different parameterizations of the production amplitude P(m) which describes formation of the $f_0(980)$ resonance in the first step
transition $e^+ e^- \rightarrow f_0(980) \gamma$ ($m$ being the $\pi^+ \pi^-$ effective mass). Let us assume that the full reaction amplitude A(m) is schematically written as a product A(m)= P(m) $\cdot$ D(m), where D(m) is the resonant $f_0(980) \rightarrow \pi^+ \pi^-$ decay amplitude. If the fits to data are performed with two different 
production functions P(m) then the fitted parameters of the resonant amplitude D(m) could be different as well. 
This is one possible source of the model dependence found in experimental analyses. 

Let us enumerate some other problems frequently encountered in the data analyses of the reactions in which the scalar mesons are produced:

1. application of the Breit-Wigner formulae with constant widths of scalar mesons,

2. reduction of a number of scalar resonances to one (for example, assuming that near the $K \bar K$ threshold
only the $f_0(980)$ resonance is present),

3. too simplified treatment of the final-state meson-meson interactions,

4. existence of two closed thresholds $K^+ K^-$ and $K^0 \bar K^0$.

\subsection{Motivation to study the the  $e^+ e^- \rightarrow K^+ K^- \gamma$ reaction}
\label{subsec-1}

Below we give some additional arguments in favour of further studies on the $e^+ e^- \rightarrow K^+ K^- \gamma$ reaction.
The branching fraction of the $\Phi(1020)$ meson decay into $\pi^+ \pi^- \gamma$ has been measured
\cite{CMD2} but the branching fraction for the $\Phi(1020) \rightarrow K^+ K^- \gamma$ channel is yet
unknown.
There are data for the radiative decay of the $\Phi(1020)$ meson into two scalar resonances.
The corresponding ratios of the widths are: 
$\Gamma(f_0(980)\gamma)/\Gamma_{total}=(3.22\pm0.19)\cdot 10^{-4}$ and 
$\Gamma(a_0(980)\gamma)/\Gamma_{total}=(7.6\pm0.6)\cdot 10^{-5}$ \cite{PDG}.
Since both scalar resonances decay into $K^+ K^-$ mesons, one should experimentally observe the reaction $e^+ e^- \rightarrow K^+ K^- \gamma$. For the decay $\Phi \rightarrow K^0 \bar K^0 \gamma$
only the upper limit $1.9\cdot 10^{-8}$ is known \cite{KLOE2009}.
In the final state of the transition process 
$e^+ e^- \rightarrow \Phi  \rightarrow K^+ K^- \gamma$ the $K^+$ and $K^-$ mesons can interact so a new information about the $K^+ K^-$ strong interactions could be obtained from data. Thus the proposed experimental analysis could provide us with potentially interesting results.

\section{Reaction mechanisms}
\label{sec-2}

\begin{figure}[h]
\centering
\includegraphics[width=13cm]{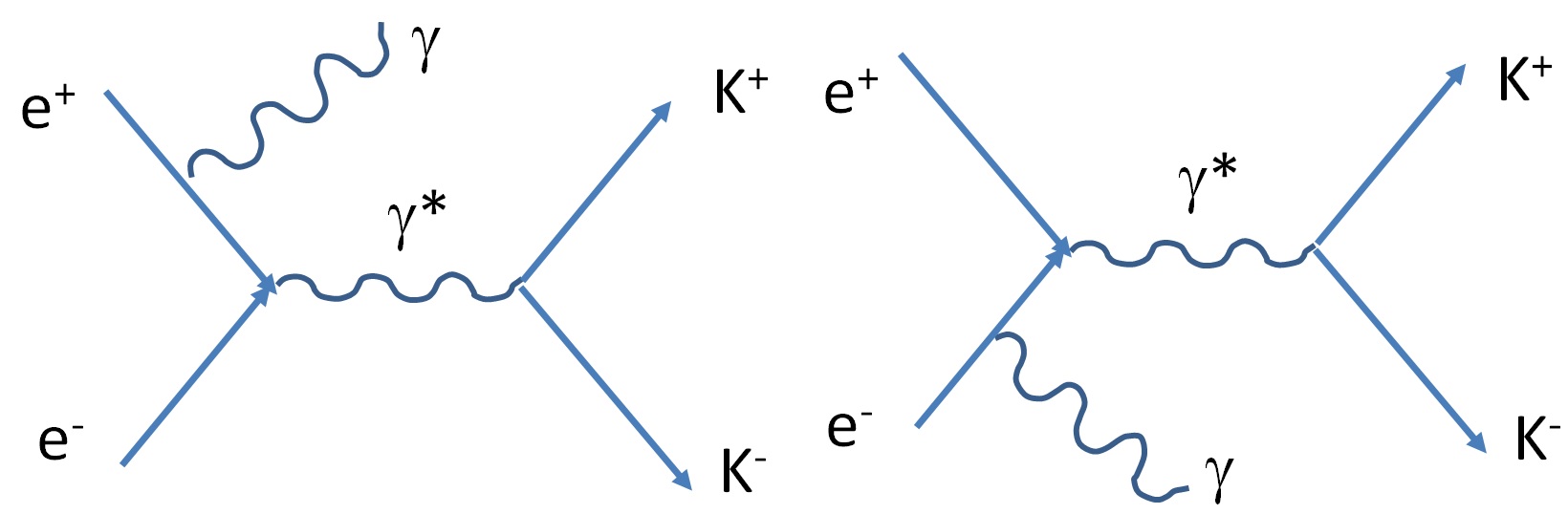}
\caption{Diagrams corresponding to the initial state radiation in the $e^+ e^- \rightarrow K^+ K^- \gamma$
reaction}
\label{fig-1}       
\end{figure}

There are several transition mechanisms processes which can lead to the same $K^+ K^- \gamma$ final state.
In the first one, called the initial state radiation (ISR) a photon is emitted from an electron or a positron
in the $e^+ e^-$ collision associated with the production of strange mesons $K^+$ and $K^-$. As shown in Fig.~\ref{fig-1} the two mesons are emitted from an intermediate photon $\gamma^*$.

\begin{figure}[h]
\centering
\includegraphics[width=13cm]{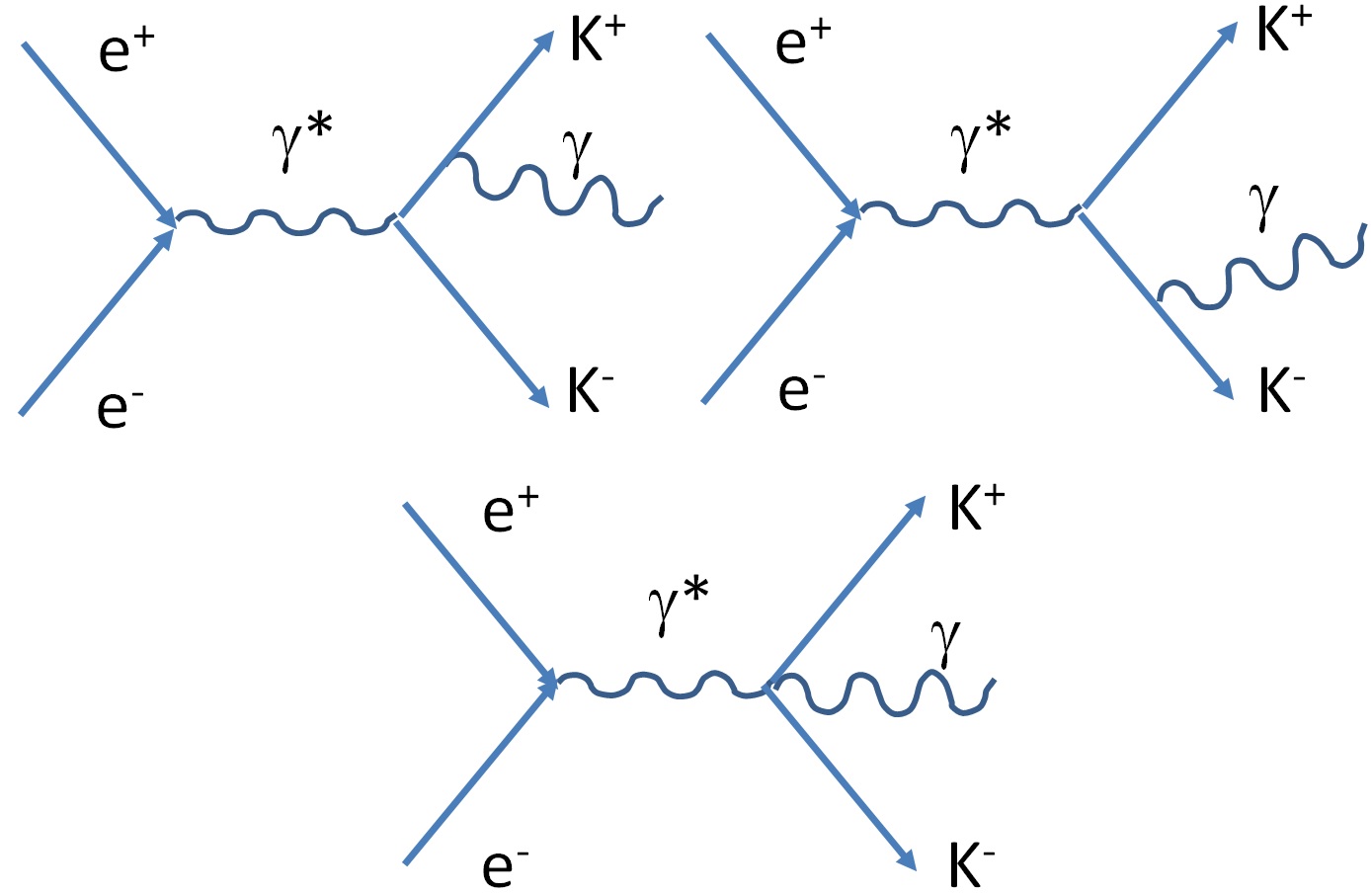}
\caption{Diagrams contributing to final state radiation in the $e^+ e^- \rightarrow K^+ K^- \gamma$
reaction}
\label{fig-2}       
\end{figure}

In other process, called the final state radiation (FSR), the outgoing photon is emitted from the $K^+$ or from the $K^-$ meson, or directly from the same vertex connecting $K^+$ and $K^-$ with $\gamma^*$. In Fig.~\ref{fig-2} the third diagram of the FSR process is called the contact term. 
Both the ISR and FSR amplitudes can be calculated using the methods of quantum electrodynamics.

\begin{figure}[h]
\centering
\includegraphics[width=13cm]{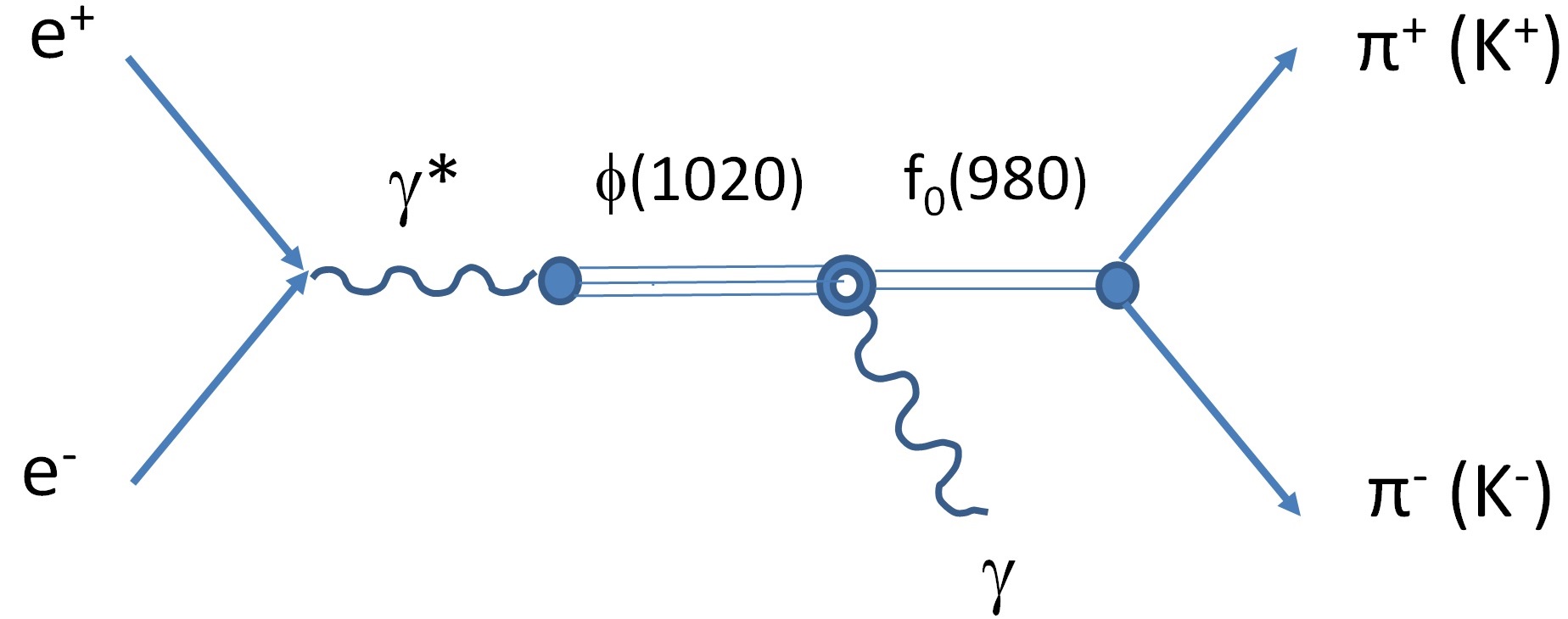}
\caption{Diagram that corresponds to the no-structure model \cite{NS}}
\label{fig-3}       
\end{figure}

The mesons $K^+$ and $K^-$ can interact strongly and in a case where the $K^+ K^-$ effective mass is
close to 1 GeV this interaction has a resonant character. Two scalar resonances can be formed:
isoscalar $f_0(980)$ and isovector $a_0(980)$. The $f_0(980)$ resonance can decay into $\pi^+$
and $\pi^-$ mesons. For a description of the reaction $e^+ e^- \rightarrow \pi^+ \pi^- \gamma$
the no-structure model (NS) has been formulated by Isidori, Maiani, Nicolaci and Pacetti
\cite{NS}. In this model an essential role is played by the $\Phi(1020)$ meson point-like coupling to 
$f_0(980)\gamma$ (see Fig.~\ref{fig-3}).
The model can be extended to describe the $e^+ e^- \rightarrow K^+ K^- \gamma$ reaction since
the $f_0(980)$ resonance decays also to the $K^+ K^-$ state.

Another model of the radiative $\Phi(1020)$ meson decays with a formation of kaon pairs has been developed by
Achasov, Gubin and Shevchenko in Ref.~\cite{KL}. In this model one calculates the amplitudes related to three diagrams shown in Fig.~\ref{fig-4}. In each of these diagrams the  $\Phi$ meson is coupled to the scalar
mesons $f_0(980)$ or $a_0(980)$ through the charged kaon-loop (KL model). Both no-structure and kaon-loop models have been used in analysis of the KLOE Collaboration for the $e^+ e^- \rightarrow \pi^+ \pi^- \gamma$ reaction \cite{KLOE2006}.

\begin{figure}[h]
\centering
\caption{Diagrams of the kaon-loop model \cite{KL}}
\label{fig-4}       
\end{figure}

\subsection{Reaction amplitudes}
\label{subsec-2} 

We have just discussed a few reaction mechanisms leading to the same  final state of $K^+K^-\gamma$.
Therefore  the corresponding amplitudes have to be added in the total reaction amplitude $M$:
\begin{equation}
\label{M}
M=A(ISR)+A(FSR)+A(KL),
\end{equation}
where $A(ISR)$ is the initial state radiation amplitude, $A(FSR)$ is the final state radiation amplitude
and as an example we have added the amplitude $A(KL)$ describing the final state interaction between kaons in the kaon-loop approach. 
The modulus of the total amplitude squared reads:
\begin{eqnarray}
\label{|M|2}
\nonumber
|M|^2=|A(ISR)|^2+|A(FSR)|^2+|A(KL)|^2+2 Re[A^*(ISR)~A(FSR)]\\
+2 Re[A^*(ISR)~A(KL)]+2 Re[A^*(FSR)~A(KL)].
\end{eqnarray}
Therefore the total differential cross-section can be witten as a sum of six terms:
\begin{equation}
\label{dsig6}
d\sigma(total)=d\sigma(ISR)+d\sigma(FSR)+d\sigma(KL)+Int(ISR-FSR)+Int(ISR-KL)+Int(FSR-KL).
\end{equation}
The first three terms are the cross-sections proportional to the moduli of the amplitudes squared
like $d\sigma(ISR)\propto|A(ISR)|^2$. 
The three inteference terms are also present, for example, Int(FSR-KL) is the term
corresponding to the interference of the FSR amplitude with the kaon-loop model amplitude.
Let us notice that if the experimental cuts are chosen symmetrically with respect to an  interchange of
the $K^+$ and $K^-$ mesons, then the two interference terms vanish: Int(ISR-FSR)=Int(ISR-KL)=0.

\subsection{Differential cross-cection}
\label{subsec-3}

Let us consider the differential cross-section of the reaction
$e^+(p_{e^+}) e^-(p_{e^-}) \rightarrow K^+(p_{K^+}) K^-(p_{K^-}) \gamma(q)$,
where the particle four momenta are indicated by $p$ or $q$:

\begin{equation}
d\sigma=\frac{(2\pi)^4}{2 \sqrt{s(s-4 m_e^2)}} |M|^2 d\Phi_3.
\label{dsig}
\end{equation}
In the formula above $s$ is the Mandelstam invariant $s=(p_{e^+}+p_{e^-})^2$,
$m_e$ is the electron mass, $M$ denotes the total reaction amplitude and $\Phi_3$
is the final state tree-body phase space.
Next we define the $K^+ K^-$ and $K^- \gamma$ effective masses squared
\begin{equation}
m^2=(p_{K^+}+p_{K^-})^2 ,~~~~~~~ m^2_{K^- \gamma}=(p_{K^-}+ q)^2
\label{mm}
\end{equation}
and two momentum transfers squared
\begin{equation}
t=(p_{e^-}-q)^2, ~~~~~~~ t_1=(p_{e^-}- p_{K^-})^2.
\label{tt1}
\end{equation}
Then the four-fold differential cross-section can be written as
\begin{equation}
\label{dsig4}
\frac{d\sigma}{dm^2 dm^2_{K^-{\gamma}}dtdt_1}=\frac{|M|^2}{(2\pi)^4 16 s (s-4 m^2_e) (s-m^2)r(t_1)}.
\end{equation}
Here 
\begin{equation}
\label{r}
r(t_1)=\sqrt{-(t_1-t_{1min})(t_1-t_{1max})},
\end{equation}
where $t_{1min}$ and $t_{1max}$ are the lower and upper limits of $t_1$ which depend on $m$, 
$m_{K^- \gamma}$ and $t$ in the following way:
\begin{equation}
\label{t1minmax}
t_{1min}=r_0+r_1 t - b \sqrt{-t(t+s-m^2)}, ~~~t_{1max}=r_0+r_1 t + b \sqrt{-t(t+s-m^2)}.
\end{equation}
The coefficients $r_0$, $r_1$ and $b$ are expressed as
\begin{equation}
\label{r0r1b}
r_0=m_K^2-\frac{1}{2}s(1-v~ z), ~~~ r_1=-\frac{1}{2}+\frac{(s+m^2)~v~ z}{2 (s-m^2)}, ~~~
b=\frac{m\sqrt{s}~v \sqrt{1-z^2}}{s-m^2}.
\end{equation}
In the above equations $m_K$ is the charged kaon mass, $v=\sqrt{1 - 4 m_K^2/m^2}$ is the $K^-$ velocity in the $K^+ K^-$ center-of-mass frame and $z$ is the cosine of the angle between $K^-$ momentum and the photon momentum in the same frame. In Eq.~(\ref{r0r1b}) we have neglected a small value of the electron mass squared.
The variable $z$ is directly related to the $K^-\gamma$ effective mass:
\begin{equation}
\label{mkg}
m_{K^-\gamma}^2=m_K^2+\frac{1}{2}(s-m^2)(1-v~z).
\end{equation}

\section{$K^+K^-$ effective mass distributions}
\label{sec-3}

\begin{figure}[htp]
\includegraphics[scale = 0.28]{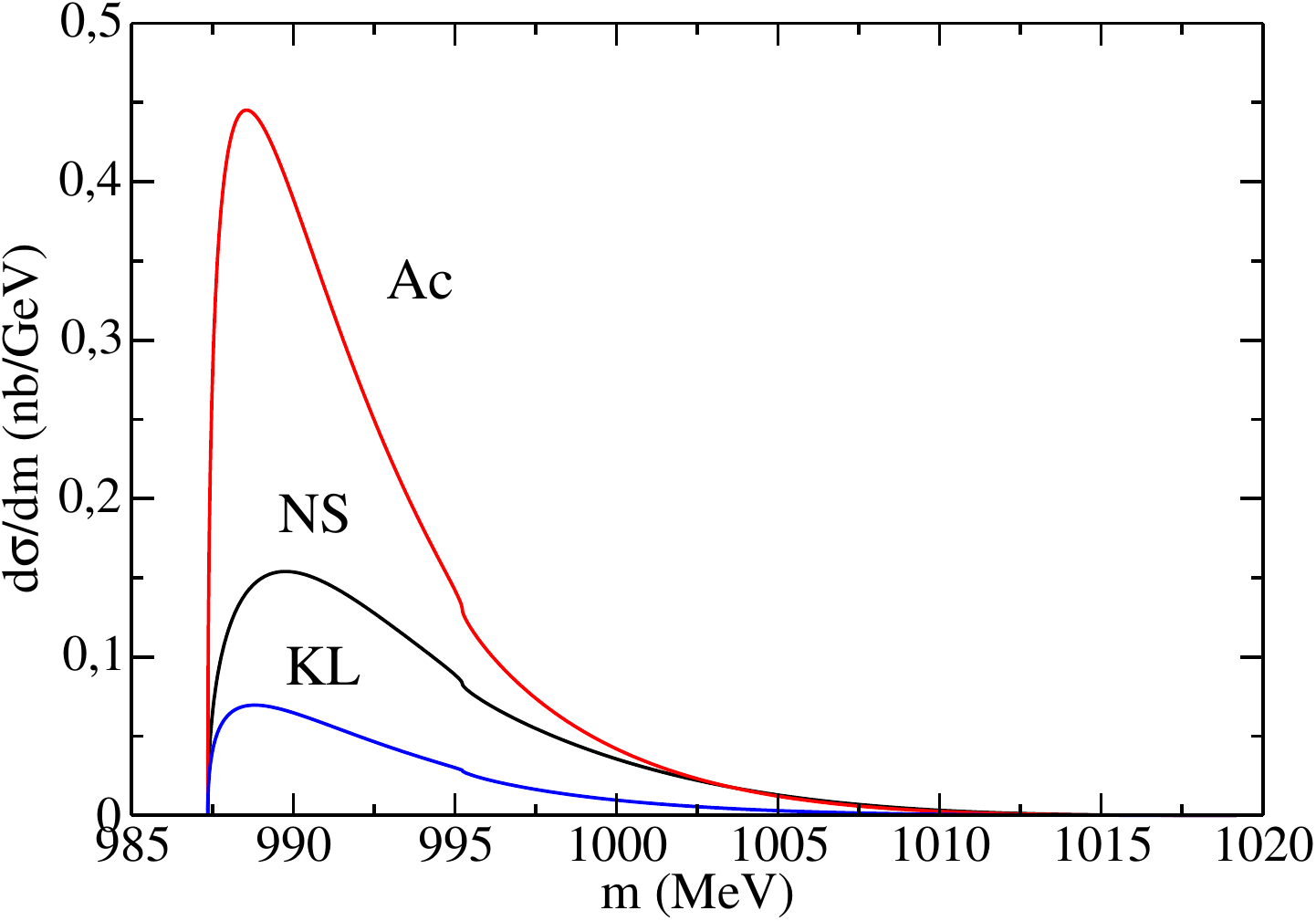}
\centering
\caption{Comparison of the $K^+K^-$ effective mass distributions calculated using three models with parameters taken from Refs.~\cite{KL} (line labelled Ac) and  ~\cite{KLOE2006} (labels NS and KL).
}
\label{fig-5}       
\end{figure}

As mentioned in the previous section, the KLOE data for the $e^+ e^- \rightarrow \pi^+ \pi^- \gamma$ process  ~\cite{KLOE2006} have been described using the no-structure \cite{NS} and the kaon-loop models \cite{KL}. 
We have extended these models and calculated the $K^+K^-$ effective mass distributions in the 
$e^+ e^- \rightarrow K^+ K^- \gamma$ reaction. The distributions we are going to discuss now correspond to the third term 
written in Eq.~(\ref{|M|2}) and denoted for the KL model as $|A(KL)|^2$. The parameters of the kaon-loop model (KL) and the no-structure model (NS) have been taken from Table 1 of Ref.~\cite{KLOE2006}. In this reference 
only the $K^+ K^-$ coupling to one scalar resonance $f_0(980)$ is present. We have also done calculations using  
somewhat different parameters of the two scalar resonances $f_0(980)$ and $a_0(980)$ as used earlier in Ref.~\cite{KL}.
In this case the masses of the $f_0(980)$ and $a_0(980)$ were equal to 980 MeV and the  coupling constants to the $K^+ K^-$ system were equal.  The corresponding results are shown in Fig.~\ref{fig-5}. 
The differential cross-section calculated for the model of Achasov, Gubin and Shevchenko is larger than the two other cross-sections. This can be simply explained by the fact that 
adding coherently two equal amplitudes related to the $f_0(980)$ and $a_0(980)$ resonances with the same parameters leads to an enhancement by a factor four in comparison with models that have only one scalar resonance.

\begin{figure}[htp]
\includegraphics[scale = 0.28]{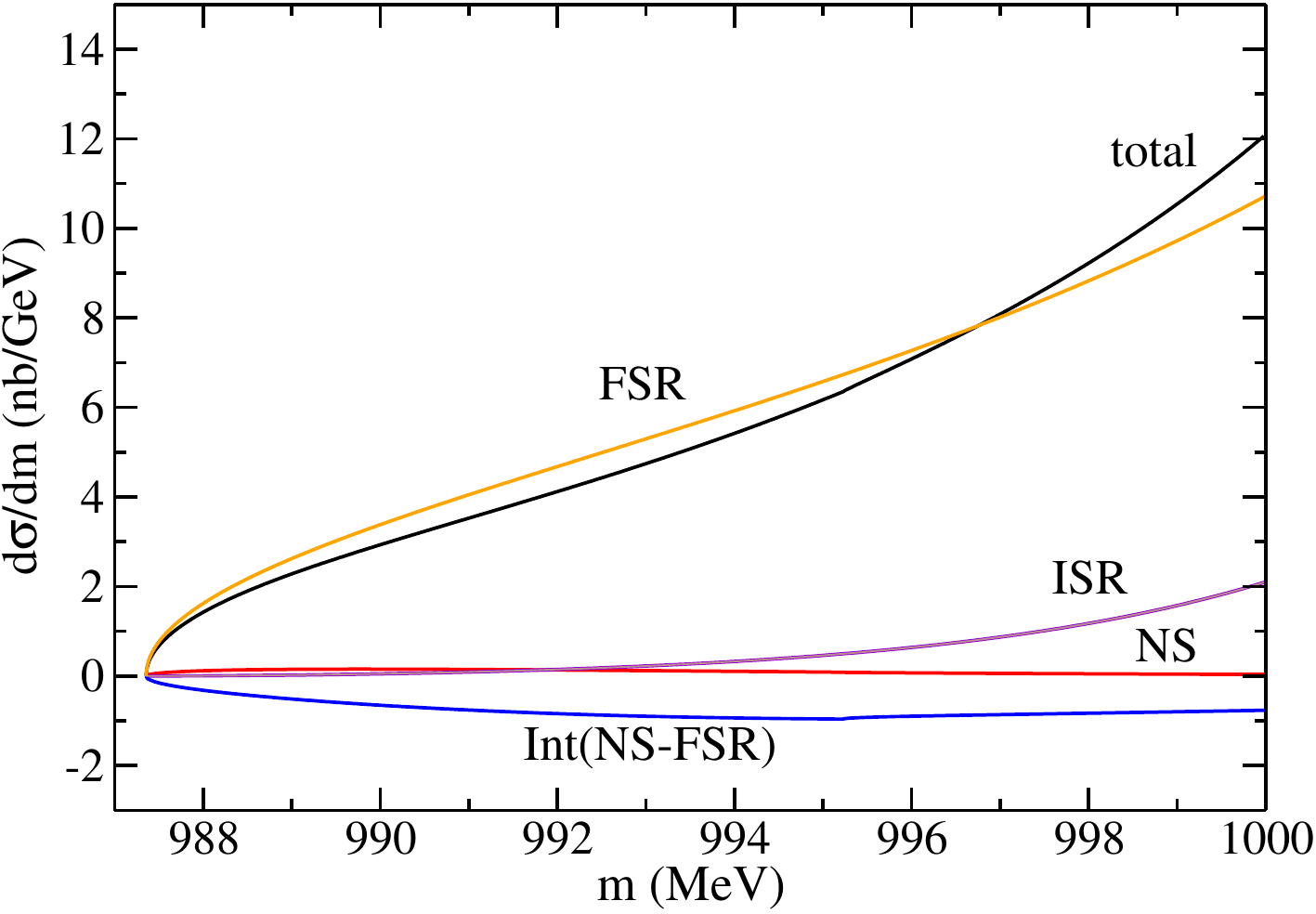}
\centering
\caption{$K^+K^-$ effective mass distributions calculated for $45^0 < \theta_{\gamma} < 135^0.$
Four contributions to the total result are labelled FSR (final state radiation), ISR (initial state radiation), NS (NS model) and Int(NS-FSR) (interference of the NS and the FSR amplitudes).
}
\label{fig-6}      
\end{figure}

\begin{figure}[htp]
\includegraphics[scale = 0.28]{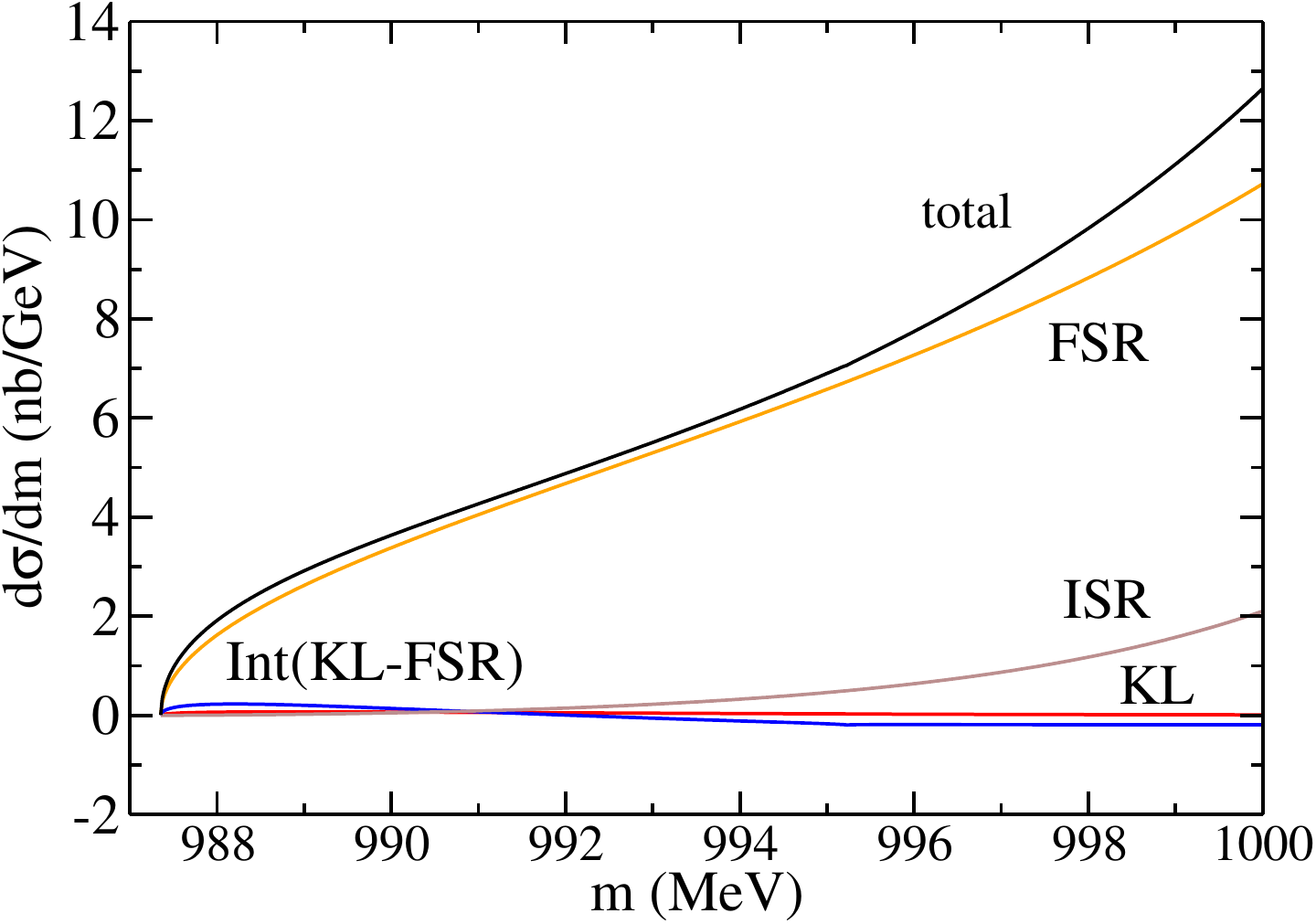}
\centering
\caption{ $K^+K^-$ effective mass distributions with $45^0 < \theta_{\gamma} < 135^0.$
Four contributions to the total result are labelled FSR (final state radiation), ISR (initial state radiation), KL (kaon-loop model) and Int(KL-FSR) (interference of the KL and the FSR amplitudes).
}
\label{fig-7}      
\end{figure}

Now  let us discuss other terms giving non-zero contributions to the differential cross-section
$d\sigma/dm$. Here we have chosen symmetric cuts on the photon emission angle $\theta_{\gamma}$ in the centre-of-mass $e^+e^-$ frame ($45^0 < \theta_{\gamma} < 135^0$). At the beginning let us take the NS model 
and make a comparison of its $K^+K^-$ effective mass distribution with the ISR and FSR ones. Apart of the 
NS, ISR and FSR distributions we show in Fig.~\ref{fig-6} lines of the  NS-FSR interference term and the total
differential cross-section. One sees that in the range of the effective mass $m$ limited to 1000 MeV 
the FSR contribution dominates while the ISR cross-section is largely suppressed by the cuts put on the angle $\theta_{\gamma}$. The NS contribution is quite small but the interference term of the NS model amplitude with the FSR amplitude is not negligible in comparison with the FSR one which gives some hope to be measured in experiment. Similar results are shown in Fig.~\ref{fig-7} for the kaon-loop model with parameters fixed in Ref.~\cite{KLOE2006}. It is interesting to see the negative contribution of the interference term in a part of the spectrum where $m$ is larger than about 993 MeV.

\section{$K^-$ angular distributions}
\label{sec-4}

\begin{figure}[ht] 
\centering
\includegraphics[scale = 0.28]{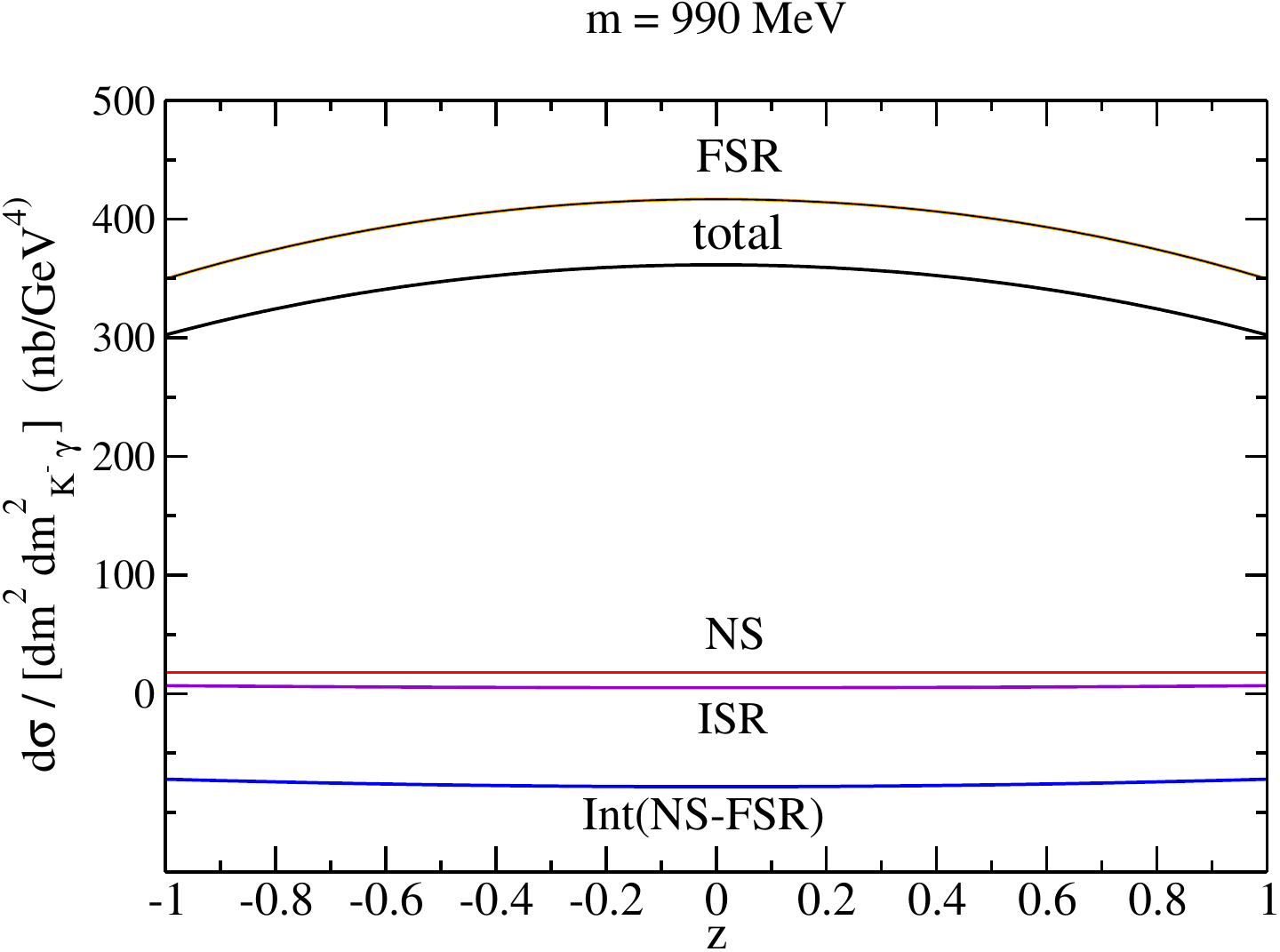}~~~~
\includegraphics[scale = 0.28]{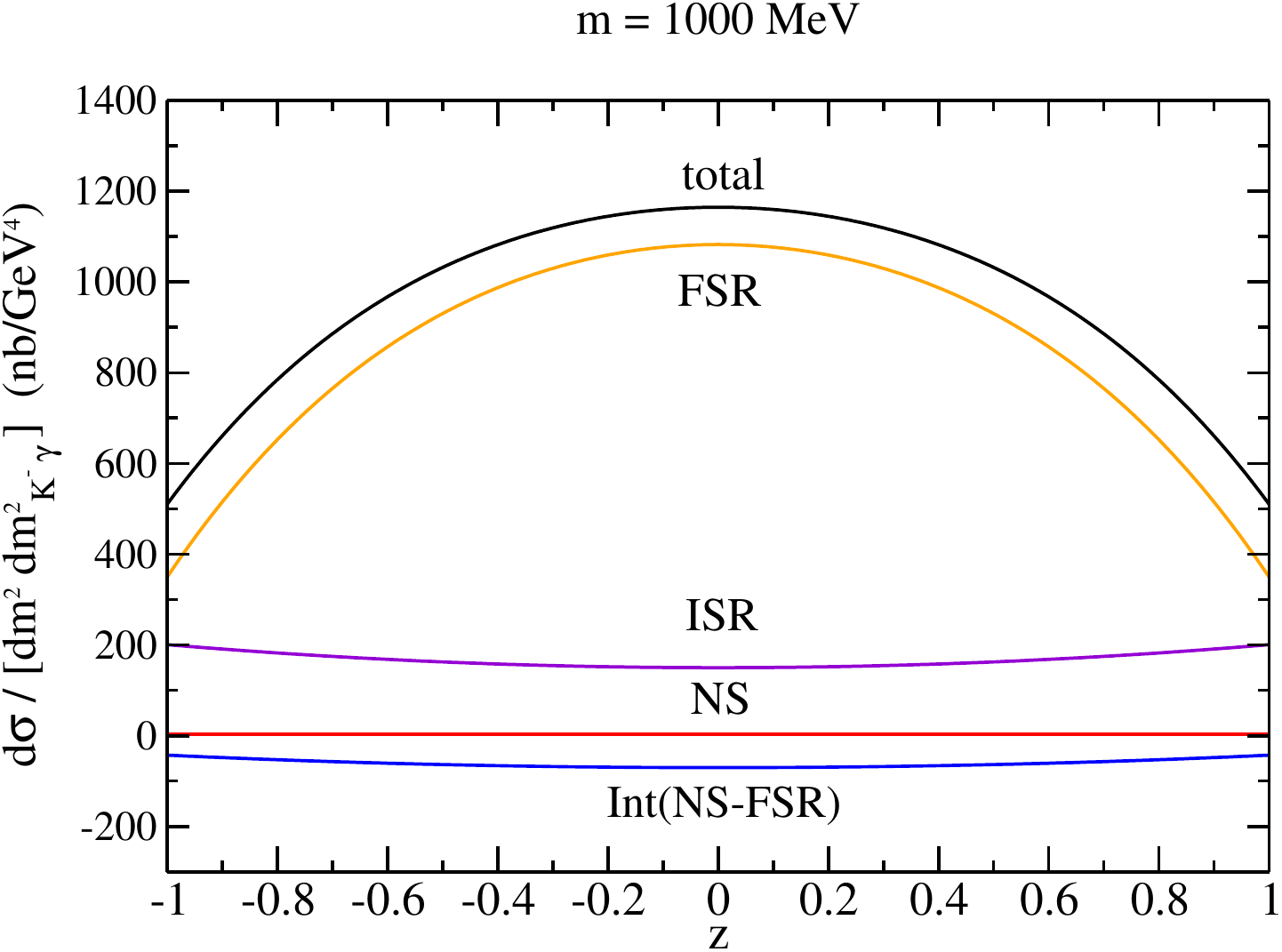}~~~~
\caption{Distributions of the polar angle of $K^-$ with respect to the photon momentum in the $K^+K^-$ center-of-mass frame at fixed $m$ equal to 990 and 1000 MeV. The curve labelled FSR corresponds to the final state radiation,
the line labelled ISR to the initial state radiation, the line NS to the no-structure model,
the interference term is labelled Int(NS-FSR), and the line labelled $total$ is a sum of all the contributions.}
\label{fig-8}      
\end{figure}

In this section we first pass to a study of the double differential cross-section
$d\sigma/[dm^2 dm^2_{K^- \gamma}]$ at fixed $m$. As seen in Eq.~(\ref{mkg}) the variation
of the $K^-\gamma$ effective mass $m$ while keeping the $K^+K^-$ effective mass fixed, is equivalent to 
a variation of $z$ which is the cosine of the angle between $K^-$ momentum and the photon momentum in the $K^+K^-$ center-of-mass frame. The $z$ distributions for two $m$ values are presented in Fig.~\ref{fig-8}. Here one sees
maxima at $z=0$ for the FSR and minima for the interference terms. The ISR function has a minimum at $z=0$ for the $\theta_{\gamma}$ range between $45^0$ and $135^0$. The distribution (NS) of the no-structure model 
is flat. This feature is common to all the models in which the $K^+$ and $K^-$ mesons interact in the $S$-wave.

\begin{figure}[ht]
\centering
\includegraphics[scale = 0.28]{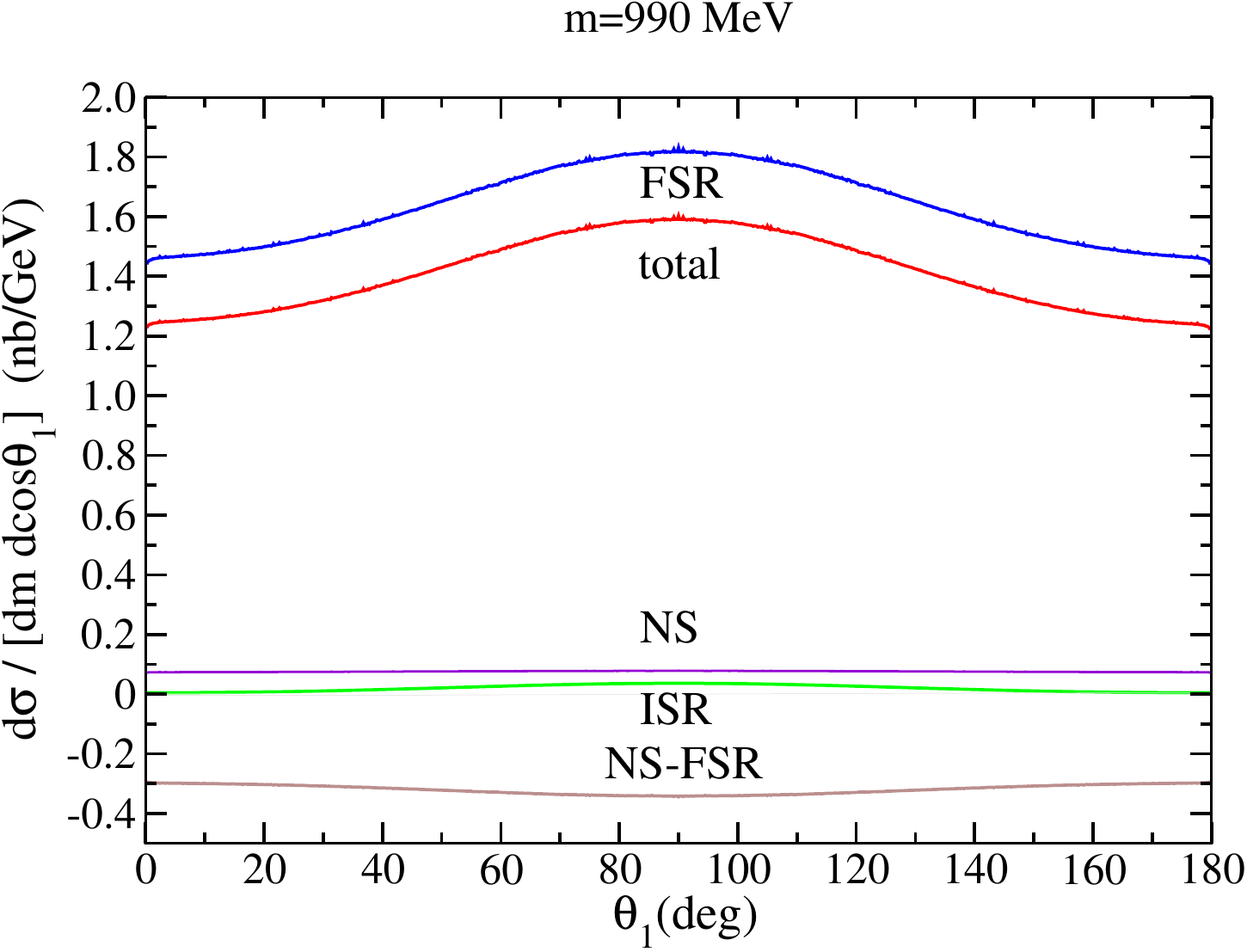}~~~~
\includegraphics[scale = 0.28]{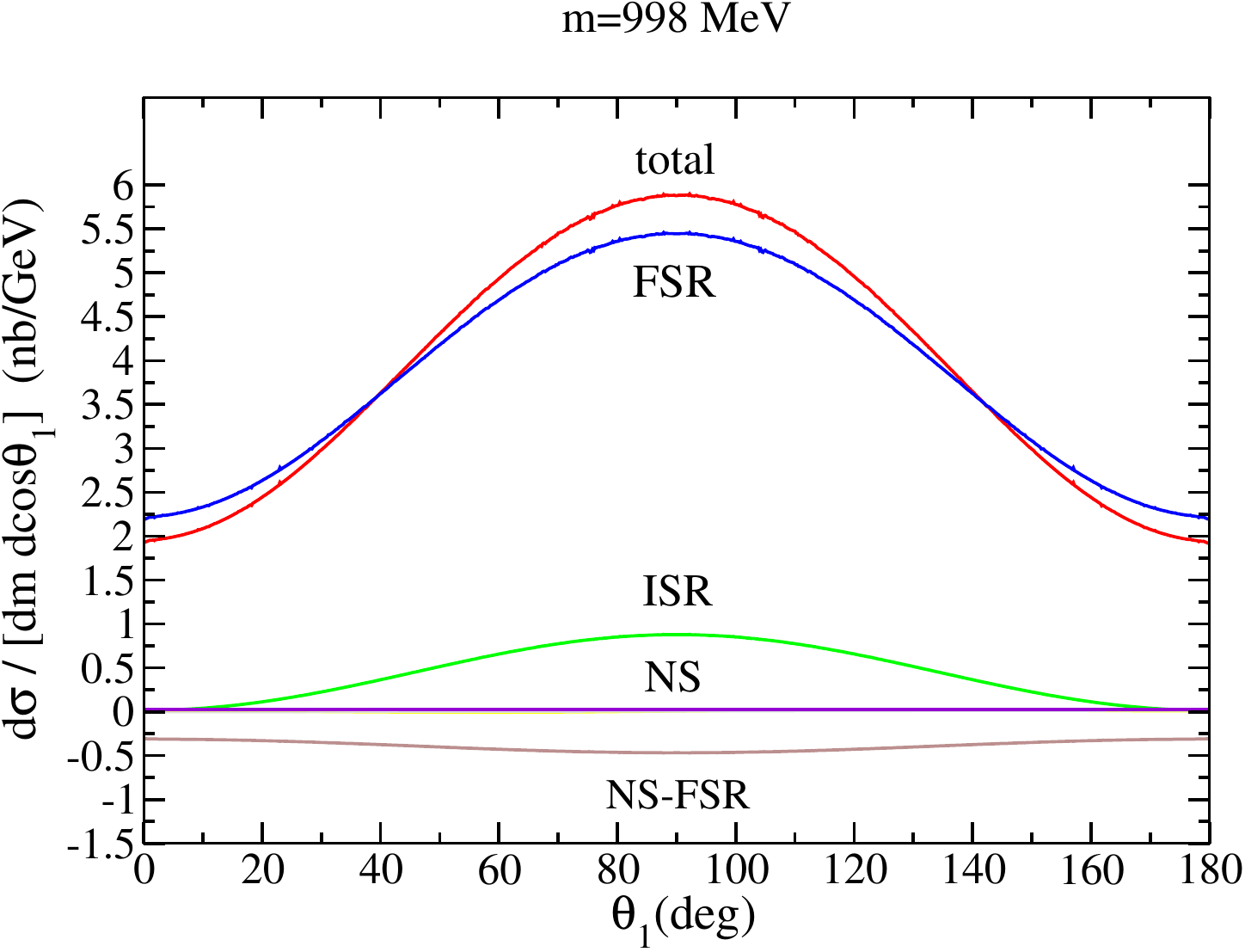}~~~~\\
\includegraphics[scale = 0.275]{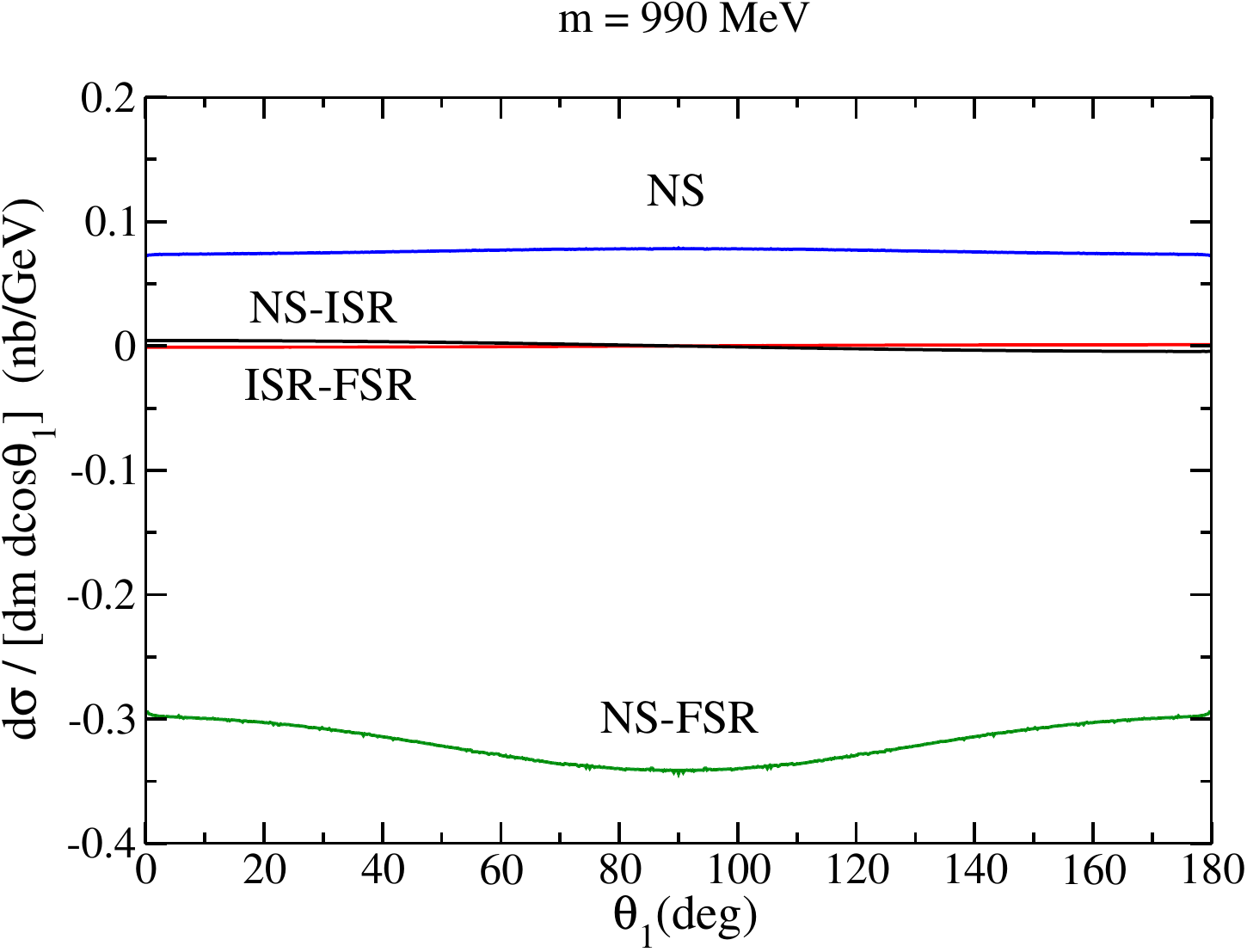}~~~~
\includegraphics[scale = 0.275]{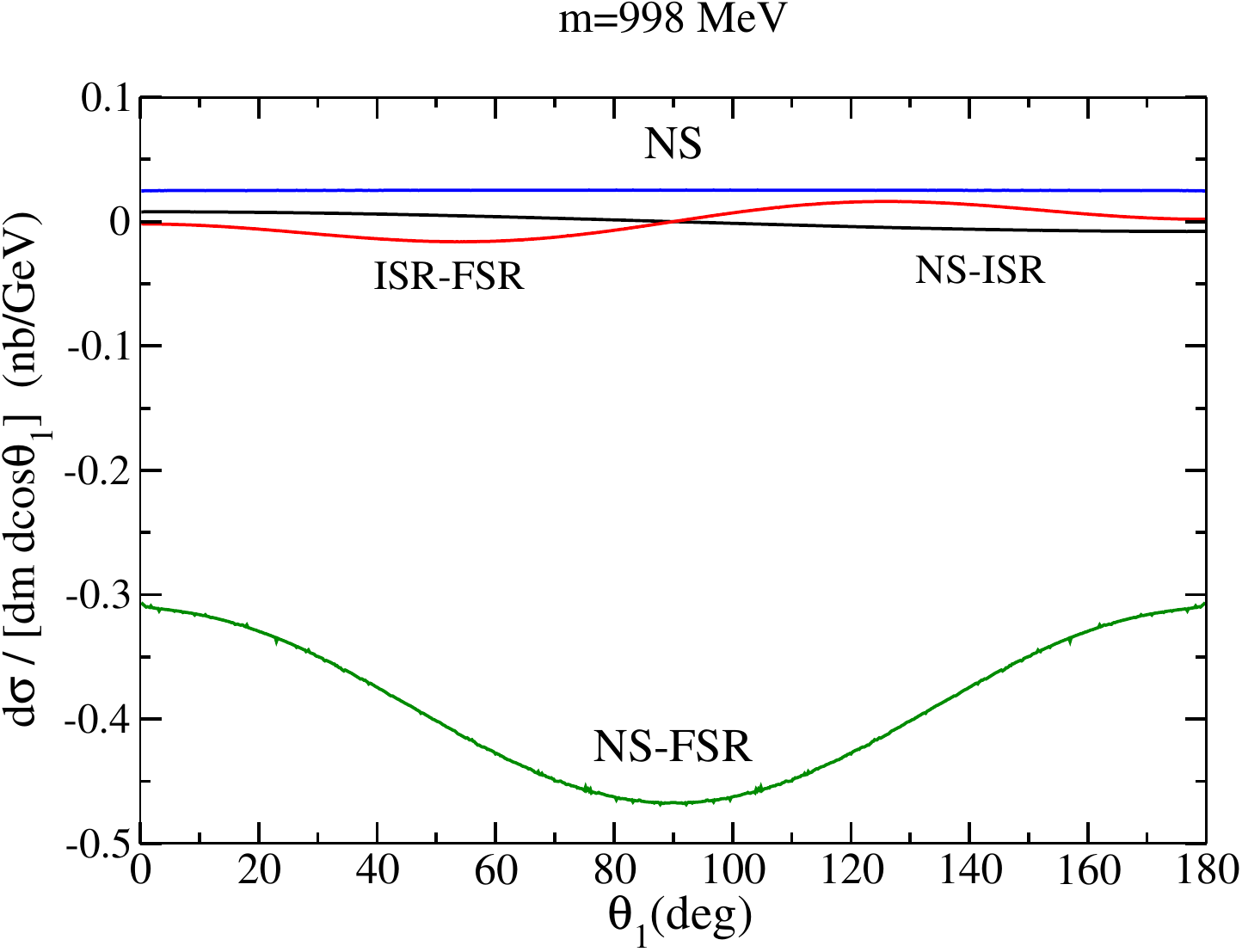}~~~~
\caption{$K^-$ angular distributions at fixed $m$ for $45^0 < \theta_{\gamma} < 135^0$. The meaning of the labels total, FSR, ISR and NS is the same as in Fig. 8.
NS-FSR, ISR-FSR and NS-ISR denote the interference terms of the NS and FSR amplitudes,
the ISR and FSR amplitudes, and the NS and ISR amplitudes, respectively. 
}
\label{fig-9}      
\end{figure}

Angular distributions of the $K^-$ mesons with respect to the electron momentum in the $e^-e^+$ center-of-mass frame have also been studied. The relevant $K^-$ angle is denoted by $\theta_1$. Once again using the no-structure model we have calculated all the six contributions to the double differential cross-section
$d\sigma/[dm~dcos\theta_1]$ seen in Eq.~(\ref{dsig6}). So in Fig.~\ref{fig-9} we notice two small interference terms ISR-FSR and NS-ISR in addition to other five lines labelled similarly as in Fig.~\ref{fig-8}. 
These two terms are asymmetric as they change sign when the angle $\theta_1$ is changed into $180^0 - \theta_1$.
Therefore they vanish after the integration over the full range of $\theta_1$. 
Let us also notice that the shape of curves changes when one increases the $K^+K^-$ effective mass from 990 MeV to 998 MeV.

\section{Photon angular distributions}
\label{sec-5}

\begin{figure}[ht] 
\centering
\includegraphics[scale = 0.28]{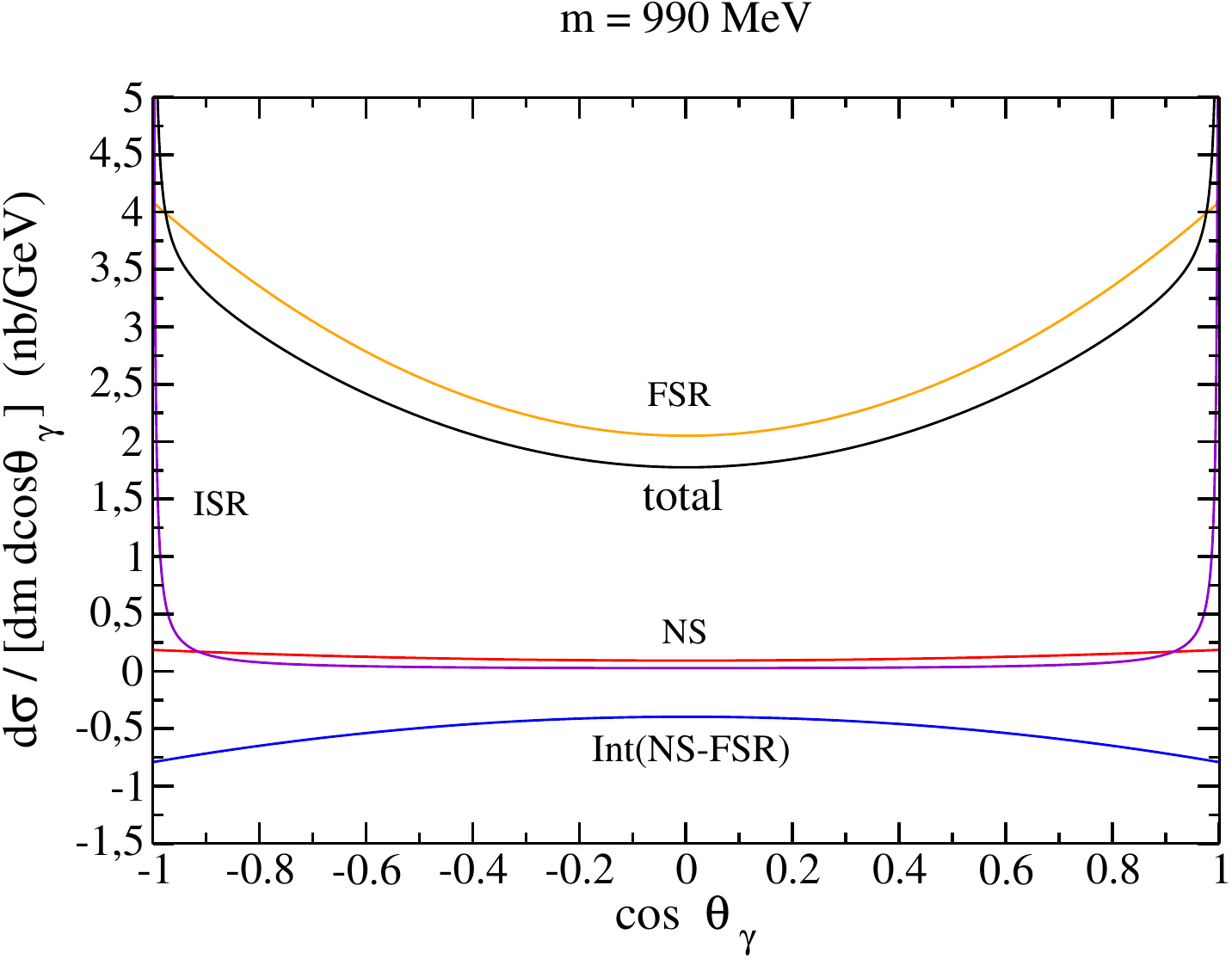}~~~~
\includegraphics[scale = 0.28]{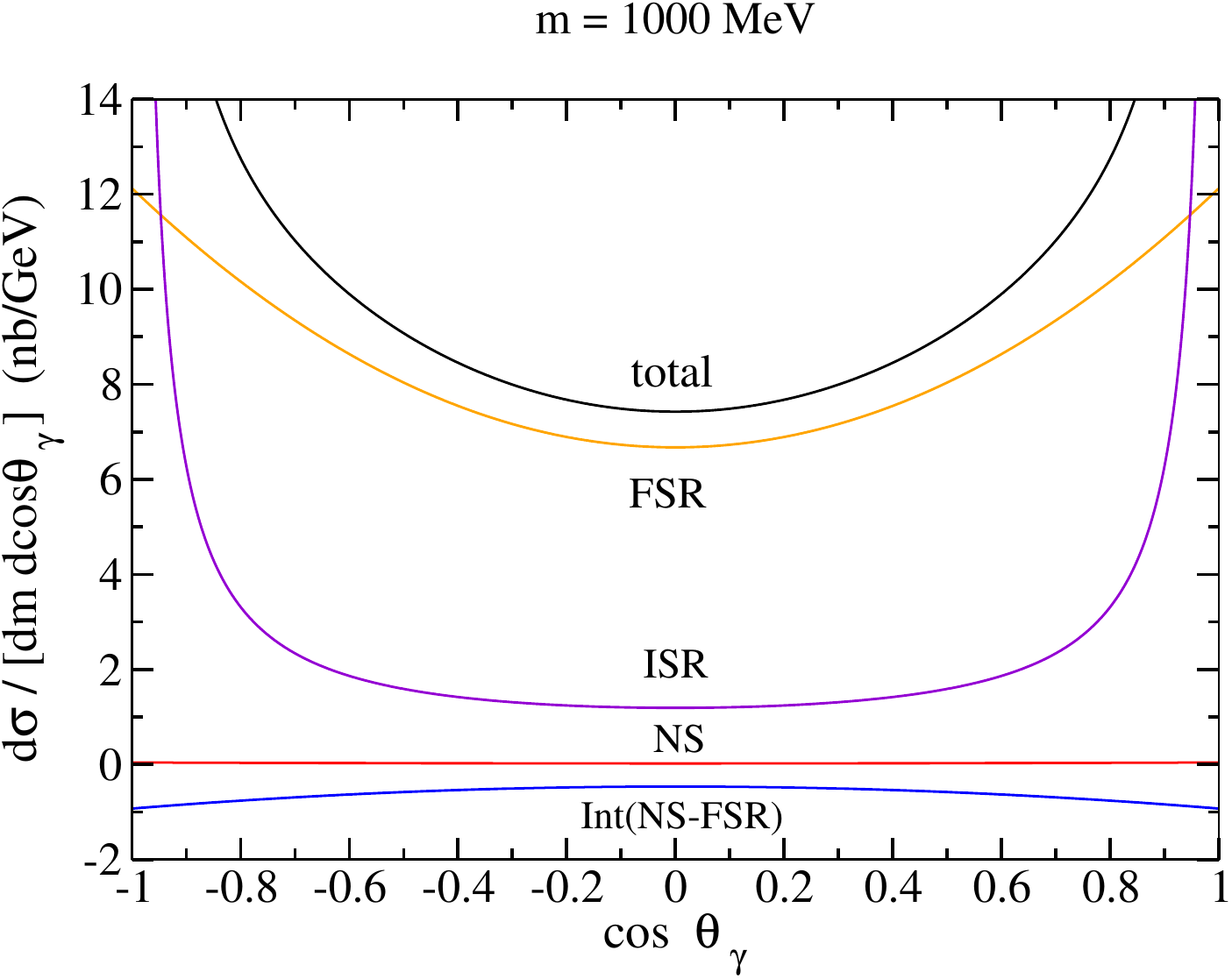}~~~~
\caption{Photon angular distributions at fixed m=990 and 1000 MeV.
For a description of lines see 
Fig. 8 caption.}
\label{fig-10}      
\end{figure}

Finally we consider the photon angular distributions with respect to the $e^-$ momentum in the $e^-e^+$ center-of-mass frame. We calculate the double differential cross-section $d\sigma/[dm~dcos~\theta_{\gamma}]$ at fixed values of $m$.
As previously we use the no-structure model. Fig.~\ref{fig-10} shows the five lines for two masses $m=990$ MeV and
$m=1000$ MeV. Its labelling is that as in Fig.~\ref{fig-8} caption. The most spectacular behaviour is a rise of the ISR term when $cos~\theta_{\gamma}$ approaches to 1 or to -1. At these two values the ISR cross-section goes to infinity and therefore in experimental studies the cuts on the values of $cos~\theta_{\gamma}$ are put
in order to diminish the ISR background. For example, in Figs. 6-8 we have cut the range
of the photon angles from 0$^0$ to 45$^0$ and from 135$^0$ to 180$^0$.

\section{Multichannel model of the $e^+ e^- \rightarrow M_1 M_2 \gamma$ reactions}
\label{sec-6}
In this chapter we briefly outline basic properties of a special model which can be
formulated for a simultaneous description of the reactions
$e^+ e^- \rightarrow M_1 M_2 \gamma$ in which $M_1$ and $M_2$ are the pseudoscalar mesons
coupled to the $K^+ K^-$ mesons. 
Here one can enumerate the following set of reactions:\\
1. $e^+ e^- \rightarrow \pi^+ \pi^- \gamma$,\\
2. $e^+ e^- \rightarrow \pi^0 \pi^0 \gamma$,\\
3  $e^+ e^- \rightarrow \pi^0 \eta \gamma$,\\
4. $e^+ e^- \rightarrow K^0_S K^0_S \gamma$,\\
5. $e^+ e^- \rightarrow K^+ K^- \gamma$.

\begin{figure}[h]
\centering
\includegraphics[width=13cm]{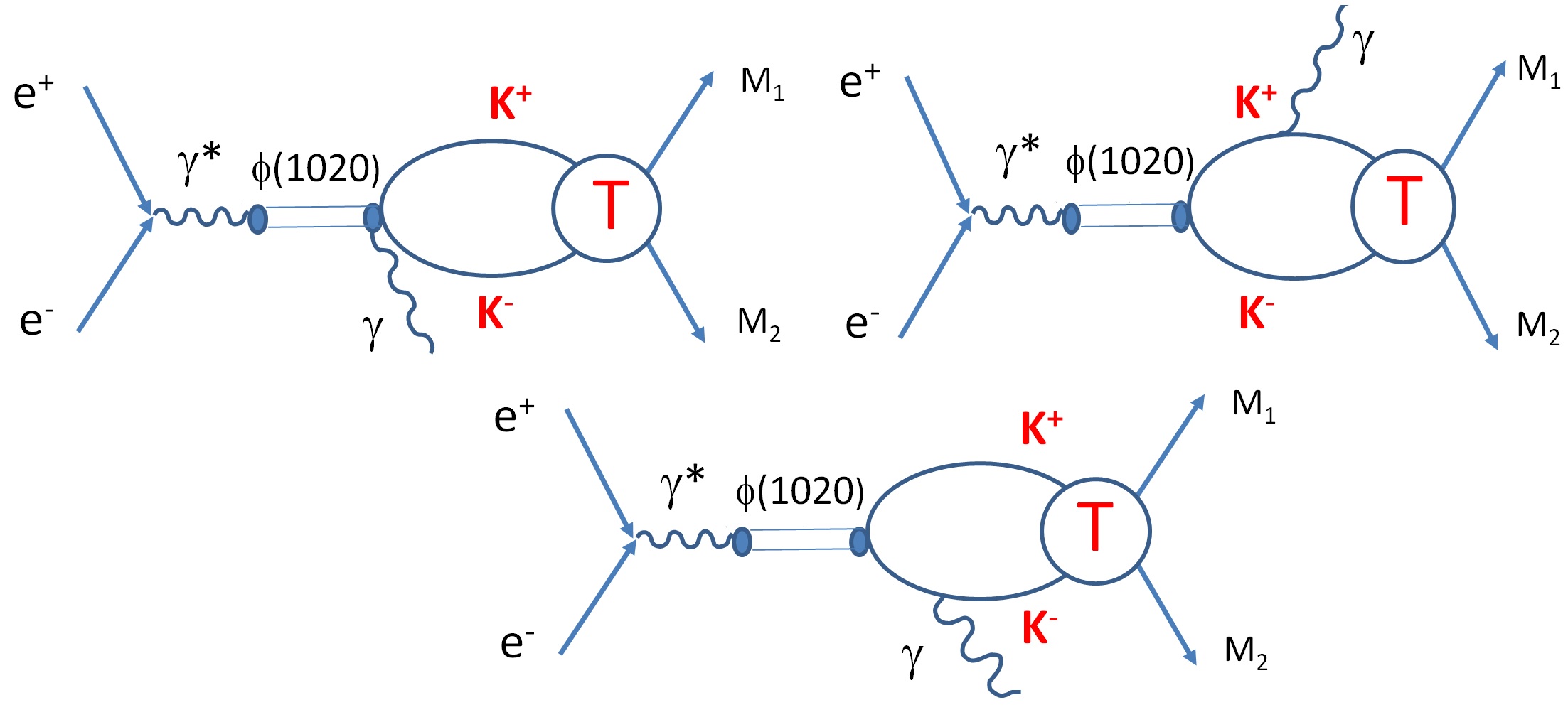}
\caption{Diagrams corresponding to the proposed model of the reactions 
$e^+ e^- \rightarrow M_1 M_2 \gamma$}
\label{fig-11}       
\end{figure}

The diagrams corresponding to the proposed model are shown in Fig. \ref{fig-11}. 
In the first step of the production process two charged kaons $K^+$ and $K^-$
and a photon are created. 
Then in the second reaction stage the kaons interact forming a sytem of two mesons $M_1$ and $M_2$. 
We denote by $T$ the corresponding set of transition amplitudes.
A specific model of these $K^+K^-\rightarrow M_1 M_2$ transitions should be unitary.
Its desirable feature is analyticity of the transition amplitudes which relates different
coupled channels. 
For a practical future description of the data 
all the transition amplitudes should have the same poles corresponding to the relevant scalar mesons
present in the energy range close to 1 GeV.

\section{Experimental implications}
\label{sec-7}

The differential cross-sections which have been calculated for different reaction mechanisms leading to the same
final state $K^+ K^- \gamma$ can be integrated within some experimental limits put on the photon
minimum energy and on the photon polar angle defined with respect to the electron beam axis.
If the minimum photon energy is equal to 10 MeV in the $e^+e^-$ center-of-mass frame then the maximum efective $K^+K^-$ mass is close to 1009 MeV. 
In Table~\ref{tab-1} the reaction cross sections are given.
Abbreviations for the reaction mechanisms are the same as in Fig.~\ref{fig-6} caption.

\begin{table}[h]
\centering
\caption{The cross sections integrated over the $K^+K^-$ effective mass from the threshold up to 1009 MeV 
for two ranges of the photon emission angle $\theta_{\gamma}$. 
}
\label{tab-1}       
\begin{tabular}{lll}
\hline
reaction mechanism & $24^0 < \theta_{\gamma} < 156^0$ & $45^0 < \theta_{\gamma} < 135^0 $\\\hline
FSR                &  0.330 nb                        &    0.238 nb\\
NS                 & 0.0020 nb                        &    0.0014 nb\\
Int(NS-FSR)        &-0.021  nb                        &   -0.015 nb\\
ISR                & 0.183  nb                        &    0.104 nb\\\hline
total              & 0.494  nb                        &    0.328 nb\\\hline
\end{tabular}
\end{table}

Assuming integrated luminosity of 1.7 fb$^{-1}$ one can obtain expected numbers of events.
In Table~\ref{tab-2} we show values of the expected number of events obtained for two different
cuts on the photon angles $\theta_{\gamma}$ and for the minimum photon energy of 10 MeV in the $e^+e^-$ 
center-of-mass frame. 
These numbers are not yet corrected for the experimental efficiency.

\begin{table}[h]
\centering
\caption{Numbers of events 
for the $K^+K^-$ effective mass up to 1009 MeV and for two ranges of $\theta_{\gamma}$
for two photon angle ranges.
}
\label{tab-2}       
\begin{tabular}{lll}
\hline
reaction mechanism & $24^0 < \theta_{\gamma} < 156^0$ & $45^0 < \theta_{\gamma} < 135^0 $\\\hline
FSR                & 5.6$\cdot  10^5$                        &   4.0$\cdot  10^5$ \\
NS                 & 3.4$\cdot  10^3$                        &   2.4$\cdot  10^3$ \\
Int(NS-FSR)        &-3.6$\cdot  10^4$                        &  -2.5$\cdot  10^4$ \\
ISR                & 3.1$\cdot  10^5$                        &   1.8$\cdot  10^5$ \\\hline
total              & 8.4$\cdot  10^5$                        &   5.6$\cdot  10^5$ \\\hline
\end{tabular}
\end{table}

\section{Summary}
\label{sec-8}

Three theoretical models have been extended in order to make predictions for the reaction
$e^+ e^- \rightarrow K^+ K^- \gamma$. 
The resulting effective mass and angular distributions can be 
used in future experimental data analyses. 
Some features of the new model of the multichannel coupled reactions $e^+ e^- \rightarrow M_1 M_2 \gamma$, where $M_1$ and $M_2$ are pseudoscalar mesons, have been outlined.
This model can be applied in a combined analysis of the radiative $\Phi(1020)$ resonance decays into two
mesons. 
It can also serve in determination of the threshold parameters of the $K^+ K^-$ strong interaction amplitudes as well as in a better specification of the properties of the scalar meson resonances $f_0(980)$ and $a_0(980)$.

\vspace{1cm}

\begin{acknowledgement}
This work has been supported by the Polish National Science Centre (grant no 2013/11/B/ST2/04245).
\end{acknowledgement}

\vspace{1cm}

%
%

\end{document}